\documentclass[10pt,leqno]{amsart}
\usepackage{graphicx}
\usepackage{indentfirst,csquotes}

\usepackage{amssymb,amsthm,amsmath}
\usepackage{xcolor,paralist,hyperref,fancyhdr,etoolbox}
\usepackage{gensymb}
\usepackage{biblatex}

\hypersetup{ colorlinks=true, linkcolor=black, filecolor=black, urlcolor=black }

\begin{document}
\title{Telepresence Lantern \\ Designing an Immersive Video-Mediated Communication Device for Older Adults} 
\author{Thomas H. Weisswange$^{1}$, Joel B. Schwartz$^{2}$, Aaron J. Horowitz$^{2}$, Jens Schm\"udderich$^{1}$}

\thanks{ $^{1}$Honda Research Institute Europe GmbH {\tt\smallskip \{thomas.weisswange, jens.schmuedderich\}@honda-ri.de}, $^{2}$Sproutel, Inc. {\tt\smallskip \{joel, aaron\}@sproutel.com}}

\maketitle


\begin{abstract}
We present the “Telepresence Lantern” concept, developed to provide opportunities for older adults to stay in contact with remote family and friends. It provides a new approach to video-mediated communication, designed to facilitate natural and ambient interactions with simplified call setup. Video communication is an established way to enhance social connectedness, but traditional approaches create a high friction to frequent connection due to, for example, technological barriers. Through interactive sessions with older adult users, we created design and function prototypes to suit their needs and preferences. The main features of our design are a curved, wide field-of-view screen and corresponding camera and sound setup, and the affordance to easily move the device from room-to-room.  An interactive user session with a fully functional prototype validated the potential of this concept for improving communication among older adults and their families.  
\end{abstract} 

\bigskip

\section{Introduction}
Socialization is a key element of living and aging happily and healthily. However, nearly half of the population aged 65 and older is living in either low-density or rural locations \cite{Ambe2019}. This often provides a challenge to frequent social interaction with family and friends that might live far away. Tee et al. \cite{Tee2009} interviewed remote families on their communication behavior and found that many people would like to communicate more but feel like they need a concrete reason to initiate a communication and they try to take care to only share “relevant” information. But based on the responses they received, the authors also saw an opportunity to enhance communication through technology.  Yuan et al \cite{Yuan2016} interviewed older adults in the US and found that in-person communication is by far the most preferred form but interestingly, for all types of mediated communication (phone, video, email) they received a significant number of negative responses, often due to limitations in “the users’ ability to experience what the medium offers”.
Beyond the “classical” communication means, there is a large number of HCI research on how to mitigate social connections or “connectedness” between distant partners. Hassenzahl et al. \cite{Hassenzahl2012} provide an extensive overview of different approaches and highlight the importance of interactions beyond exchanging information. Consistently, many proposals in the community can be considered “phatic technologies” \cite{Vetere2009} that provide means to enhance “ambient awareness” \cite{Gaver2002}. An example specifically considering older users is the Messaging Kettle \cite{Brereton2015}, designed based on research about emotionally meaningful everyday objects for older adults, which allowed sharing abstract and textual messages around a joint activity to overcome physical and social distances. There exist also multiple prototypes around interactive picture frames \cite{Beacker2014,Mynatt2001} which allow older adults to stay connected to their families. In general, more and more approaches try to involve older adults directly in the design of technical systems to better cover their specific needs and abilities \cite{Ambe2019,Eisma2004,Fischer2020}.
Despite the positive effects of such technologies, they are usually not meant to replace directed communication but provide an additional means to enhance connectedness \cite{Howard2006} and as Vetere et al wrote “phatic interactions are not limited to being peripheral” \cite{Vetere2005}. Several studies showed that providing social contact to relatives through video-mediated communication (VMC) has the potential to improve quality of life and decrease loneliness for older adults in nursing homes \cite{Bennett2015,VanderHeide2012,Mickus2002,Savolainen2008,Tsai2020}. 
However, it is also noted that the quality and reciprocity of such communication is an important factor, and, if it is low, can even increase loneliness \cite{Karimi2012}. Compared to communication over telephone, VMC is considered as creating more closeness \cite{Ames2010,Brubaker2012,Kirk2010} and sense of presence \cite{Savolainen2008}. Kirk et al \cite{Kirk2010} found that many people use it to be more involved in others live and even sometimes to partake in daily routines. An interesting aspect they found was also that people liked moving the camera around for example to show the remote party a certain object or place. In a study on long-distance relationships \cite{Neustaedter2012}, participants also reported to move their VMC device between different rooms, despite non-supportive designs, to be able to share more time together.  There are also reports on the frequent use of VMC systems as continuously open channels to share time, despite being typically designed around a “calling activity” \cite{Neustaedter2015}.
VMC is also thought to be better suited for interactions with groups and with small children \cite{OHara2009}. In fact, the communication between grandparents and grandchildren is one of the dominant reasons to use VMC within families \cite{Tee2009,Ames2010}. But while older adults prefer richer context and focused communication, their grandchildren tend to have more fluid communication patterns \cite{Forghani2014,Yarosh2010} and there are good use-cases for moth face-to-face and contextual camera angles \cite{Gaver1993}. In classical desktop or laptop computer based VMC the framing and camera angle often limit the usefulness for such dynamic scenarios or calls with multiple participants on one side \cite{Kirk2010}. It is also known that narrow angled cameras and the two-dimensional video image limit free interactions due to missing joint reference frames \cite{Gaver1993}. 
In general, many older adults are still reluctant to use VMC due to complex and difficult-to-use technology and lack of dedicated designs \cite{Beacker2014,Eisma2004,Yuan2016}. Ames et al \cite{Ames2010} also found that setting up and scheduling of VMC is considered a large effort for both grandparents and their remote families and hinders more frequent interaction. 

The work presented in this paper will propose a VMC device developed in interaction with the target user group to be more approachable and accessible for the older population. Additionally, we will address a second objective by proposing and testing features that aim to improve quality and foster reciprocity of video communication. In the next section we introduce prior work focusing on either of these two objectives and afterwards describe our design approach to provide solutions to the combined problem. We define explicit design objectives and opportunities and explain our concept of a “Telepresence Lantern”. Combining user-centered and creative design approaches (see e.g. \cite{VettingWolf2006}) we explore form factors and in parallel assess technological feasibility, which culminates in a functioning prototype device. Key improvements include the combination of a wide camera angle, an immersive viewing experience to spatially represent the viewing angle, and a convenient and acceptable embodiment for enhanced usability. The functional prototype along with target design renderings is then tested in detailed interactive interviews with 6 older adult participants.

\section{Related Work}
\subsection{Older Adults-directed VMC design}
In previous work, designing VMC tools for older adults was approached in a variety of ways. Boman and colleagues \cite{Boman2012} used Inclusive Design to create a simple videophone mock-up, fit to the needs of people with dementia, with a focus on an accessible interface. Some of their design guidelines, which are stated to be also relevant beyond people with dementia, are a familiar and age-relevant look, minimal interface complexity and tolerance for errors. Munoz et al. \cite{Munoz2015} developed a front end for social networks and communication means including VMC that was fit to the needs and interests of older adults. Their “SocialConnector” system increases communication through an easy access for both symmetric and asymmetric communication means. Despite generally positive responses to the interface, they found that using touch and voice interactions was frequently leading to confusions. Zamir and colleagues \cite{Zamir2018} provided a “Skype on Wheels” prototype to the residents of several care environments. They used a collaborative action research approach to identify remaining barriers and then enhance the feasibility of their design for older adults. Although they gathered some positive feedback on VMC in general, both staff and older adults were hesitant towards the proposed design and handling. One reason for this was that most of the interfacing was still very technical. Addressing this issue, KOMP by No Isolation B.V. \cite{Braenden2018} is a product designed to provide a simple interface and a look resembling a familiar object for older adults - a classical tube TV. The device also offers the possibility for family members to send pictures and messages. However, calling is only possible using an app but not from the device itself, so the older person cannot initiate communication herself. In their report, the authors also describe additional similar prototypes such as an always-on “Window” or a VMC device designed to look and work similar to a radio. 

\subsection{Communication-improving VMC design}
An interesting alternative approach to adapting the interfaces of VMC applications for older adults is to provide social connections through enhancing shared experiences through VMC. Massimi \& Neustaedter \cite{Massimi2014} describe a survey on the potential of VMC for participating in family events and limitations such as issues with camera field of view and placement but also the difficulty to provide an ambient feeling but also support more intimate directed conversations. Licoppe et al \cite{Licoppe2017} discusses the problems of classical video communication for extended and contextual interaction with the example of presenting objects to a remote partner. One idea they put forward to improve such interactions was the use of 360\degree camera and screen for VMC. Family windows \cite{Judge2010a} was a set of devices that was built to improve the connectedness between remote parties through an “always-on” video connection on a tablet computer. Their field study found that participants were using it to share everyday experiences and infer contextual information related to, for example, appropriate phone call scheduling (for another work on context awareness for communication initiation, see also \cite{Riche2010}). However, most participants were missing means to directly initiate a communication with the device, as it did not provide any audio connection. In a follow up study using a commercial VMC software \cite{Neustaedter2015} that was able to have both an always-on and a calling functionality, the authors confirmed the positive effects. The main remaining issue they reported was the use of multi-function tablet computers or mobile phones, which prevented a consistent setup, the authors suggested a dedicated device to be a better solution. The idea of more continuous video connections on a dedicated device is also followed by the Portal by Facebook, Inc (https://portal.facebook.com/), which even fosters new ways of VMC through augmented reality features. HomeProxy \cite{Tang2013} provided a dedicated device for direct VMC and combined it with an activity sensor that allowed for both hands-free interaction and providing information about the presence of a person to the remote site to ease initiation of a call. Guo et al \cite{Guo2019} introduced a mobile phone app that embedded a classical face-to-face video communication system with a panoramic image of the current location to increase context awareness and provide cues for additional communication topics. Li and colleagues \cite{Li2018,Li2019} proposed a spherical display and a 360\degree camera for VMC in collaborative work settings. The feedback from their user study was very positive, in particular with respect to context awareness, and multiple participants mentioned the potential of similar technology for private communication. One limitation was the low resolution due to the used projection technique.
All those approaches showed good potential in extending the field-of-view and providing means for more casual communication setups. However, they were not tested with older adults, nor did they specifically consider this user group in the design process.
 
\section{Designing the Lantern}
\begin{figure}
    \centering
    \includegraphics{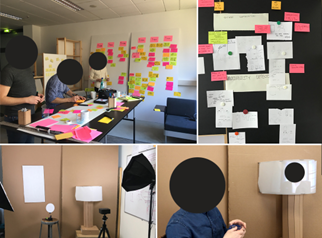}
    \caption{Workshop sessions, including concepting, ideation, and clustering (top); rapid prototyping interactions with paper mockups and a mini projector, with studio lights for video documentation (lower left); and bodystorming interactions (lower right).}
    \label{fig1}
\end{figure}

Our design process consisted of four ideation workshops with the four authors of this work, as well as two interactive user sessions to test concepts. Two of the authors come from a technology background and the other two from a background in user-centered design. Our collaboration followed a user-centered approach, focused on the needs and requirements of the target audience, and balancing usability with technological feasibility as we designed, built, and tested prototypes. Figure \ref{fig1} shows some impressions from the workshops.

\begin{figure}
    \centering
    \includegraphics{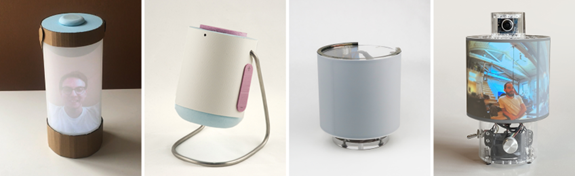}
    \caption{A paperboard mock-up of the Telepresence Lantern concept (left). A physical design prototype (center-left) and the functional prototype used in the final user tests (center-right). A fully functional technology prototype of the device (right).}
    \label{fig2}
\end{figure}

At the outset of our process, the prompt for our first workshop was much broader than VMC. We began with the question: "How might we connect generations to facilitate social connections that create happiness?". Among many ideas and concepts, we gathered our focus on two themes: increasing opportunities for connection, and the importance of joint experiences. During our second remote workshop, we generated more focused concepts and used a combination of bodystorming activities and cardboard mockups to test our ideas with rapid prototypes. We explored how we might simplify initiating VMC and augment the experience with a more approachable and less technological feeling to address how we might reduce barriers to VMC for older adults. For example: what if we could enchant the furniture of the living room with VMC? At the conclusion of this second workshop, we decided to focus on a concept that we termed “Telepresence Lantern” (Fig. \ref{fig2} far left).

The Telepresence Lantern combines video calling and 360° projection in an approachable interface with a form factor that is playful and attractive. The Telepresence Lantern is a lightweight, elegant, and portable object with functionality to make and receive video calls. The shape of a lantern, resembling existing household objects, should reduce the feeling of using a high-tech device and make it more approachable to the target audience. The form and materiality should also make the lantern more acceptable to be on prominent display in the home, for ease of use. Portability should be implied through the design and will facilitate natural and convenient communication as it allows people to set up a call in a variety of locations or even change rooms during a call. A spherical camera in the top of the lantern captures the user’s environment. This should reduce friction around setting up an appropriate framing for a call and allow more dynamic activities. A wide angle, curved display should provide additional context for a conversation, while affording spatiality to the calling experience, as opposed to viewing on a flat surface. It was previously shown that convex screens can provide high immersion while still also rating high on perceptibility \cite{Urakami2021}. The camera's wide field of view creates opportunities for intelligently zooming, cropping, or tracking participants, an area we explored in user testing. It was clear from discussion with potential users that the concept must also be trustworthy. An integral shutter or closure mechanism clearly communicates when the device is off by covering to the camera, acknowledging privacy requirements.
\begin{figure}
    \centering
    \includegraphics{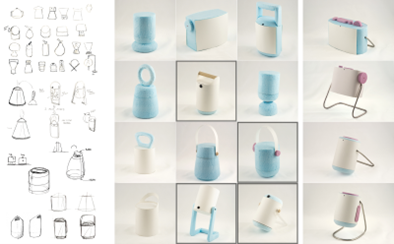}
    \caption{A selection of sketches exploring form (left), and a series of low fidelity physical design prototypes exploring shape and affordances (center). Feedback from user testing with a selection of low fidelity design prototypes (center, outlined) informed the creation of a refined set of physical design prototypes (right).}
    \label{fig3}
\end{figure}
Using these concept ideas, we began a creative design process to explore form factors through mood boards, sketches, and physical prototyping (see Fig. \ref{fig3} left and center for some examples). As we explored in 3D by bringing our sketches into form with a series of low fidelity foam models, we maintained consistent sizing to keep our focus on the varying shapes, proportions, and affordances.
We then gathered evaluative feedback on 4 of our low fidelity design prototypes (Fig. \ref{fig3} center, outlined) in interaction tests with older adult participants (n=5, ages 70-85) at a senior center in East Providence, Rhode Island, USA. The general comments on overall shape and proportions were reaffirming that our direction was desirable. We clearly saw most of the handle designs receiving interactions implying native portability (moving between spots, lifting) and related comments. Participants preferred prototypes with elevated and/or posable screens, to lift the screen and interface above the table surface, and to facilitate a natural viewing angle for VMC. With this feedback, we selected two design directions to explore further: designs with a handle that could pivot behind the device and designs with a combination handle-stand that could rotate 180° between the two functions while also raising the screen above the tabletop. To guarantee accessibility and usability, particularly among older adult users, we also established a design goal for the interface to be explicit and understandable, with tactile buttons and clear labeling. In our third workshop, we collected ideas for functionality, design, and placement of such physical interface elements. Some of the resulting interface proposals were included as physical mockups in a set of refined prototypes (Fig. \ref{fig3} right). We included more traditional, rectangular screen shapes in this prototype set as a point of comparison, to challenge the design direction of the Telepresence Lantern. We then again gathered feedback from participants representative of the adult children of older adult users (n=6, ages 26-65), at a coworking site in Providence, Rhode Island using these refined prototypes, and a technological mockup with videos played on a circular screen. We asked participants what they perceived the function of the interface components to be on these prototypes and prompted them to complete interactions such as placing a call, hanging up, or changing volume, to understand where they expected to look for this feature.  One insight about the user interface we gained from feedback was the importance of distinguishing functionality with button shape and location. We also saw that the addition of a physical "on" switch could be effective to communicate that the device is safe and secure. Results from this user test also informed key aspects of our final design: a non-rotating handle with the clear and unambiguous affordance of portability, and an elevated screen for ease of viewing. Comparing the prototypes with the context of the technological mockup, 4 of 6 the participants were more interested in form and appearance of the circular screen than in the rectangular designs, affirming our initial direction for the circular design of the Telepresence Lantern.
In our final workshop, we discussed a variety of render designs that have been made using the insights from the previous testing sessions. For the interface, we decided to include a multifunctional knob at the center that can be used to browse through contacts and change volume and if pressed initiates or accepts a call. Functionality is clarified by illumination on the knob. In addition, a physical lever clearly communicates if the device is on or off. The final design that we selected to be shown in the user interviews can be seen in Fig. \ref{fig4}. 
\begin{figure}
    \centering
    \includegraphics{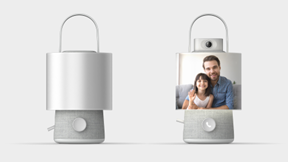}
    \caption{Rendering of final design of the Telepresence Lantern. Turned off and camera hidden (left) and active on a video call (right).}
    \label{fig4}
\end{figure}
Our design incorporates fabric, a longstanding design element of home speakers that has recently been re-adopted by electronic products to evoke warmth and familiarity, aligned with our goals to make the Telepresence Lantern approachable and less technological in feeling. Finally, to earn a place in the living rooms of our users, our interviews affirmed that it is critical to address privacy concerns. We designed the camera with a popup feature for integrated privacy, to communicate security and trust and unambiguously indicate to the user when the camera is off. 

\section{Building the Lantern}
In parallel we began building a series of technology prototypes to test the basic functionality and fit of available technical components. The final functional technology prototype was a combination of commercial and pre-commercial sample components and fully video-chat capable (Fig. \ref{fig2} center-right) in order to test it with users. We combined two flexible, touchscreen displays (7.8-inch, 1920x1440) around a 5.25-inch acrylic structure into a single, 270° display. Display driver boards were used to connect the display to an external computer (Intel NUC8i7BEK), together with a small 8W USB speaker and a 360° UVC camera with integrated microphone (Ricoh R Development Kit, 1920x960 resolution). VLC media player was used for video playback and Zoom for video chat. Prototyping with a 360° camera and very wide display curvature of 270° allowed us to evaluate the effectiveness and utility of these large angles and refine toward target specifications for our final device. Pre-recorded videos were captured with a GoPro MAX 360° camera and edited with Adobe Premiere to create stimuli with different levels of zooming and cropping for user feedback. 
We also created a self-contained, fully functional prototype (Fig. \ref{fig2} right) in the silhouette of the final design. This device was completed after the user testing discussed herein and is currently being used for ongoing testing and development.

\section{Testing the Lantern}
\subsection{Setup}
We used the technological prototype along with renderings of the final design (Fig. \ref{fig4}) to evaluate the Telepresence Lantern in detailed, IRB-approved, interactive interviews with 6 older adult participants (2 female, age 70-72). Aside from their age, inclusion criteria where having at least some experience with video chat and having a positive relationship with their family, as we hypothesized that sustained video chat is most common among family members and very close friends. Besides of general feedback, we wanted to evaluate the following research questions: 
Q1: Does a wide-angle setup improve involvement in communication and the feeling of sharing experiences as compared to prior experience of the participants (i.e. with “flat” VMC applications)?
Q2: Does a less technical design and a more implicit framing improve approachability of a VMC device?
Q3: What preferences do older adults have towards VMC, and how might this influence our design?
The sessions started with an introductory part, including an informed consent form and general questions on demographics, communication behavior (in particular on prior experience and frequency of use of VMC) and technology receptiveness. Next, we showed the Telepresence Lantern prototype and the participants used it in an interactive video call with one of the remote experimenters which was captured through a wide-angle camera mimicking another lantern on their side (Examples in Fig. \ref{fig6}). After introducing themselves, the remote experimenter prompted a brief conversation by sharing that they were planning a vacation and asking the participant if they had any favorite trips or recommendations for places to visit. During the call, the remote experimenter got up and walked around the room to do some household tasks, leaving and returning to the field of view of the device while still conversing. The experimenter also prompted the participant to get up and retrieve a prop from the wall behind them. This interaction was intended to both showcase the wide field of view of the camera and screen and provide a prompt for further discussion about feelings toward interaction quality when the remote experimenter was not on camera. This was designed to contrast classical face-to-face communication situations to see if this would disrupt the continuity or enhance the feeling of joint presence. The call took 10 minutes with another 5 minutes dedicated to questions about the experience. 

We next presented three pre-recorded videos (each about 3 minutes long) showing different activities on the Lantern to demonstrate a variety of possible shared experiences (cooking, children paying, having dinner) with another set of structured questions afterwards. This part of the interview was finalized by simulating interactions with the device and a hypothetical remote communication partner again using pre-recorded videos. 
\begin{figure}
    \centering
    \includegraphics{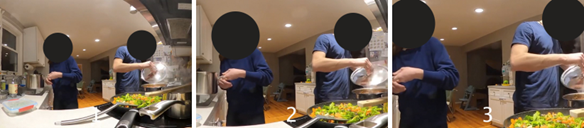}
    \caption{Example of the pre-recorded video of cooking used in the user study. Left to right shows the three different levels of zooming and cropping, used to evaluate the effect of narrowing the apparent field of view on the technology prototype's curved screen. Videos are approximately 24:9 aspect ratio; screenshots are cropped horizontally to fit this page.}
    \label{fig5}
\end{figure}

We asked participants to stand at different distances while replaying videos with different levels of zooming and cropping (Fig. \ref{fig5}) and asking the participants to share their opinions and preferences to evaluate the feasibility and acceptability of these interactions. Using a card sorting task, we also gathered feedback to understand the desired performance of technological aspects such as video or audio quality. Six different technologies, with brief explanations, were presented on cards, and participants were asked to sort these in order of importance. These included: microphone, speaker (audio quality), visual clarity (ability to see detail in the video), visual tracking (auto- framing to keep the focus in front of you), screen size, and control (ability to control what you see in the other person's room). We concluded the interview by showing the design renderings (examples in Fig. \ref{fig4}) and questions on possible consumer interests and the experiences. Overall, the sessions lasted approximately 70 minutes.
\subsection{Results}
We will describe the results of the user sessions according to the previously introduced research questions. A video with excerpt of the interactive interviews can be found in the supplementary material. 

\subsubsection{Involvement \& Experience}
\begin{figure}
    \centering
    \includegraphics{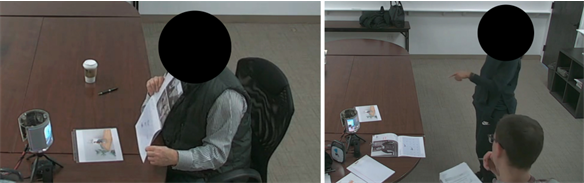}
    \caption{Examples of dynamic user interactions with the technology prototype during the interview sessions.}
    \label{fig6}
\end{figure}

During the interactive call, we saw participants interacting with the device in various dynamical ways (examples shown in Fig \ref{fig6}). 

Along those lines, they also explicitly reported that it did not interrupt their interaction when the remote experimenter was leaving the scene, just as it is natural to have a conversation while moving about a room, not always engaged in eye contact. It was also stated that the device encouraged more complex interactions, one participant for example said: "\textit{Facetime is two dimensional, if I want to show you something over there, I have to turn it and kind of aim it. Here you don’t have to do that.}"
Participants expressed that there was novelty in how immersive the experience felt, statements included "\textit{It's cool because it’s more than a box}", "\textit{It’s almost one step towards a hologram, it's almost three dimensional}" or "\textit{It has a wider range than just the computer screen or the telephone screen, so it made it seem a lot more interactive}".
After watching the experience videos, we asked questions about the feeling of being part of the activities and if participants could imagine use the device for such interactions that are less focused on direct communication. The response differed strongly between the videos, but some respondents expressed that seeing more of the remote experimenter’s environment would make it easy to prompt conversation about what they see (“[I might ask] \textit{why are you putting that spice in now}”). We also got suggestions for further experiences, that participants were considering to be interesting to share using the Telepresence Lantern, in particular various outdoor activities, like gardening or picknicks.
Interestingly, most participants indicated that they would use the Telepresence Lantern in addition to other technologies they currently use, while in some previous studies it was rather seen as a replacement for a phone call \cite{Kirk2010}. For shorter more targeted communication the telephone was still the preferred means, but seemingly there is a type of communication that our lantern good provide better than existing solutions.

\subsubsection{Approachability}
General Responses from participants were positive and enthusiastic, confirming the acceptability and likeability of the concept. Compared to other VMC applications, convenience and simplicity of our device was highlighted. When we asked participants for general feedback on the device after seeing renderings of the final design and in relation to other communication means they are using we predominantly received responses that compared the design to familiar everyday objects, like a lantern or a cooler, and explicit references to the “non-technical” look compared to tablets and laptop computers. The portability aspect of the device was mentioned by all participants ("\textit{The handle is asking you to move it}”). This was rated very positively as being unrestrictive for where to start a conversation.
In the interactive sessions, participants moved their body and got closer to the device to frame the video where it felt comfortable. None of them seemed to miss the feature of seeing their own framing as video feedback. We saw similar directions in the responses to our questionnaire for interest in automatic zoom and user tracking, where some participants expressed little interest because of their ability to "self-edit” and the feeling that it naturally mimics in-person dynamics ("\textit{You don't need to follow me over to the stove} […] \textit{or wherever else I might be going.}").
People showed interest in further functionalities, explicitly mentioning for example to asynchronously share videos and pictures through the device (as previously proposed by others \cite{Heshmat2017,Tang2013}).

\subsubsection{VMC Insights}
All participants reported general technology interest and had significant prior experience with video-mediated communication, but they also reported relying on others for technical help, particularly around initial setup. We found that convenience and low friction are paramount: people seem to choose video chat services based on what the people calling them use or the phone they own. 
The results of the card sorting unanimously showed that quality of the audio signal was the most important aspect for a successful video-mediated communication, particularly compared to the importance of the video quality (similar to e.g. \cite{Kirk2010}). This result is in accordance with positive participant reports on interaction quality for scenes when the remote partner left the camera field of view and with the frequent opinion that auto-framing does not add much to such quality. The device screen size is approximately the size of a tablet; all participants found it to be appropriate. When testing different zoom levels and interaction distances, we found that communication was acceptable up to 3m distance to the device and standing, while most people preferred a distance below one meter for any direct interactions and people preferred zoom levels that preserved sufficient context of the room rather than focusing the full screen directly on their correspondent's face.
In the responses to the rendered images, we also confirmed our assumption that people would like to start a conversation at different locations, while the typical use cases were not including carrying the Telepresence Lantern while moving around.

\section{Conclusion}
Staying connected with remote-living friends and family is an important topic of our time. Video-mediated communication has the potential to enhance social connectedness and increase interactions. However, existing solutions were designed for a younger user group. For older adults, there is a high friction to frequent connection in traditional video conferencing due to technological barriers for setting up a call, the desire of older adults to not disrupt lives of immediate family and a felt need for concrete communication reasons and topics. 
We created the Telepresence Lantern concept, which has the potential to address these problems by providing a simple and easy to understand interface, a furniture aesthetic and opening the opportunity to simply spend time together through ambient connection. The current paper provides first evidence, that some of the design decisions, that we have taken, incorporating frequent involvement of potential users, have the potential to improve both approachability of video-mediated communication technology for older adults and the experience of talking to remote friends and families. Using design and function prototypes our interactive user sessions could trigger very positive reactions to the non-technical design and reduced interface, which could imply possibly higher adoption rates. The participants liked the effect of the wide-angle camera and screen design that provided an immersive communication experience. The reported results are based on intensive interviews with a small sample of target users and interactions took place in a laboratory setting, so future work will need to include field testing of the device with more natural communication set-ups to provide more evidence for the above insights, as well as additional research of long-term effects with respect to enhancing connectedness of older adults.
Our goals of inclusive design and early testing were challenged by the technological nature of VMC and the novelty of a circular screen. Our challenge was how to conduct user testing early, to inform and verify our design direction before – and while –investing in the development of more functional prototypes. Despite the inherent challenges, we think we found a successful balance and reaffirmed that limitations of early testing were offset by the usefulness of talking to participants, which also ultimately helped inform subsequent participant interviews and our internal design workshops in very meaningful ways. One learning was the importance of tangible, non-abstract stimuli. Using multiple pre-recorded videos to test zooming and tracking, for example, was an efficient way to gather participant response, rather than building the technological feature and attempting to recreate the scene (e.g. cooking) during the interviews. 
In summary, we have presented a new concept for video-mediated communication that has the potential for a positive impact on connectedness between older adults and the remote relatives. We could see positive responses by a first group of test users confirming our approach to involve older adults and their perspectives early and often in the design process. We hope our work can also provide meaningful insights to other researchers in this space. 

\printbibliography

\end{document}